\documentclass[vecphys]{svmult}

\usepackage{makeidx}         % allows index generation
\usepackage{graphicx}        % standard LaTeX graphics tool
\usepackage{multicol}        % used for the two-column index
\usepackage[bottom]{footmisc}% places footnotes at page bottom
\makeindex             % used for the subject index
\usepackage{amsmath}
\usepackage{amsfonts}

\begin{document}

\title*{Quantum Percolation in Disordered Structures}
\author{Gerald Schubert \and Holger Fehske}
\institute{Ernst-Moritz-Arndt-Universit\"at Greifswald, 
Institut f\"ur Physik, Felix-Hausdorff-Str. 6, D-17489 Greifswald, Germany
\texttt{\{schubert,fehske\}@physik.uni-greifswald.de}}

\maketitle

\section{Introduction}
Many solids, such as alloys or doped semiconductors, form crystals
consisting of two or more chemical species. In contrast to amorphous
systems they show a regular, periodic arrangement of sites but the 
different species are statistically distributed over the available 
lattice sites. This type of disorder is often called compositional.
Likewise crystal imperfections present in any real material give rise 
to substitutional disordered structures. The presence of disorder 
has drastic effects on various characteristics of physical systems,
most notably on their electronic structure and transport properties.

\begin{figure}[t]
  \centering 
  \includegraphics[width=0.95\linewidth,clip]{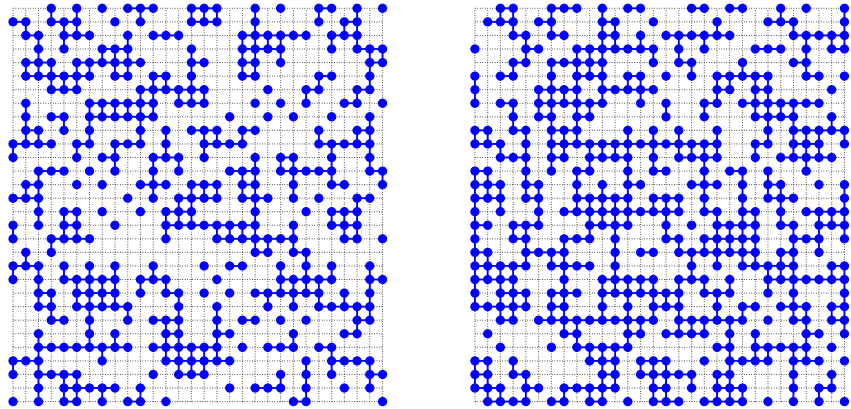}
   \caption{$2D$ site percolation. Shown are lattice realizations 
below (left) and above (right) the classical percolation threshold
$p_c=0.592746$}\label{fig_cp}
\end{figure}

A particularly interesting case is a lattice composed 
of accessible and inaccessible sites (bonds). Dilution of the 
accessible sites defines a percolation  problem for the 
lattice which can undergo a geometric phase transition between 
a connected and a disconnected phase. Since `absent' sites (bonds) 
act as infinite barriers such a model can be used to describe 
percolative particle transport through random resistor networks 
(see Fig.~\ref{fig_cp}). 
Another example is the destruction of magnetic order in diluted 
classical magnets. The central question of 
{\emph classical percolation theory} is whether the diluted lattice 
contains a cluster of accessible sites that spans the entire sample 
in the thermodynamic limit or whether it decomposes into small 
disconnected pieces. 

\begin{figure}[t]
  \centering 
  \includegraphics[width=0.95\linewidth,clip]{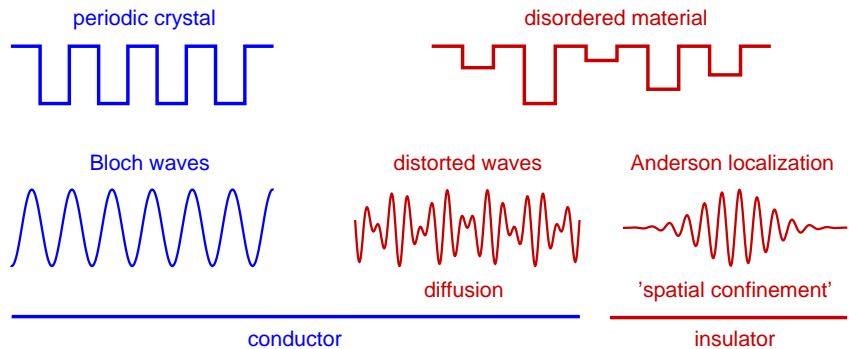}
   \caption{Anderson transition in disordered systems}\label{fig_anloc}
\end{figure}

The corresponding problem of percolation of a {\emph quantum particle}
through a random medium is much more involved. Here the famous 
concept of Anderson localization comes into play~\cite{An58}. 
Anderson argued that the one-particle wave functions in macroscopic, 
disordered quantum systems at zero temperature can be exponentially 
localized. This happens if the disorder is sufficiently strong 
or in energy regions where the density of states is 
sufficiently small~\cite{KM93_2}. 
The transition from extended to localized states is a result of 
quantum interference arising from elastic scattering of the particle 
at the randomly varying site energies. 
Particles that occupy exponentially localized states are restricted
to finite regions of space. On the other hand, particles in extended
states can escape to infinity and therefore contribute to transport 
(see Fig.~\ref{fig_anloc}).

For the quantum percolation problem all one-particle 
states are localized below the classical 
percolation threshold $p_c$, of course.
%
%However,  
%Kirkpatrick and Eggarter showed that localized states 
%always also exist on the infinite spanning 
%cluster (for all $p>p_c$)~\cite{KE72}. 
%
Furthermore for arbitrary concentrations of conducting sites $p>p_c$
there always exist localized states on the infinite spanning 
cluster~\cite{KE72}. 
Since for a completely ordered system ($p=1$) all states are extended
and no states are extended for $p<p_c$, one might expect a disorder-induced 
Anderson transition at some critical concentration 
$p_q$, with $p_c \leq p_q\leq 1$. The reason is that 
the quantum nature of particles makes it harder 
for them to pass through narrow channels, despite the fact that quantum 
particles may tunnel through classically forbidden regions~\cite{VW92}.
As yet the existence of a quantum percolation threshold, different
from the classical one, is discussed controversial, in particular
for the two-dimensional ($2D$) case where also weak localization effects
might become important~\cite{LS96}.   

Localization and quantum percolation have been the subject of 
much recent research; for a review of the more back dated
work see e.g.~\cite{Od86,LR85}. The underlying Anderson
model (with uniformly distributed local potentials) 
and site percolation or binary alloy models (with a bimodal 
distribution of the on-site energies) are the two 
standard models for disordered solids. Although both problems
have much in common, they differ in the type of disorder,
and the localization phenomena of electrons in substitutional alloys  
are found to be much richer than previously claimed. For instance,
the binary alloy model exhibits not one, as the $3D$ Anderson model,
but several pairs of mobility edges, separating localized from extended
states~\cite{KE72,SWF05,AF05}.
%
% Moreover there appears to be
%`forbidden energies', e.g. at the center of the band,
%at which the density of states goes continuously to zero.
%
Moreover it appears that `forbidden energies' exist, in the sense that near
the band center the density of states continuously goes to zero.
These effects might be observed in actual experiments. 

Understanding these issues is an important task which we
will address in this tutorial-style review by a
stochastic approach that allows for a comprehensive 
description of disorder on a microscopic scale. This approach, 
which we call the local distribution~(LD) approach~\cite{AF05,AF06},
considers the local density of states~(LDOS),
which is a quantity of primary importance in systems
with prominent local interactions or scattering. 
What makes the LD approach `non-standard' is that it directly deals with the
{\emph distribution} of the LDOS in the spirit that Anderson
introduced in his pioneering work~\cite{An58}. 
While the LDOS is just related to the amplitude of the electron's
wave function on a certain lattice site, its distribution captures the 
fluctuations of the wave function through the system.

The present paper will be organized as follows. After introducing the 
basic concepts of the LD approach in Sect.~\ref{sec:LDOS_approach} and 
explaining how to calculate the LDOS via the highly efficient
Kernel Polynomial Method~(KPM), we exemplify the LD concept by a study of 
localization within the Anderson model. Then, having all necessary tools at 
hand, we proceed in Sect.~\ref{sec:QPERC} to the problem of quantum percolation. 
Refining previous studies of localization effects in quantum
percolation, we will demonstrate the `fragmentation' of the spectrum
into extended and localized states. The existence
of a quantum percolation threshold is critically examined. 
To this end, we investigate for the $2D$ site-percolation  
problem the dynamical properties of an initially localized wave packet on 
the spanning cluster, using Chebyshev expansion in the time domain. 
In Sect.~\ref{sec:adv_mat} we apply the findings to
several classes of advanced materials where transport
is related to percolating current patterns.
Here prominent examples are $2D$ undoped graphene
and the $3D$ colossal magnetoresistive manganites. Finally, in 
Sect.~\ref{sec:conclusion} we conclude with a short summary of the topic.
\section{Local Distribution Approach}
\label{sec:LDOS_approach}
\index{local distribution approach}

\subsection{Conceptional background}

% Wichtigkeit Verteilung
In the theoretical investigation of disordered systems it turned out
that the distribution of random quantities takes the center
stage~\cite{An58,AAT73}.
Starting from a standard tight-binding Hamiltonian of noninteracting 
electrons on a $D$-dimensional hyper-cubic lattice with $N=L^D$ sites,
the effect of compositional disorder in a solid
may be described by site-dependent local potentials $\epsilon_j$.
\begin{equation}
  \label{model}
  {H} =  - t \sum\limits_{\langle ij \rangle} 
  \bigl({c}_i^{\dag} {c}_j^{} + \text{H.c.}\bigr)
  + \sum\limits_{j=1}^{N} \epsilon_j {c}_j^{\dag} {c}_j^{} \;.
\end{equation}
The operators ${c}_j^{\dag}$ (${c}_j^{}$) create (annihilate) an electron in
a Wannier state centered at site $j$, the on-site potentials $\epsilon_j$ 
are drawn from some probability distribution $p[\epsilon_j]$, and $t$ 
denotes the nearest-neighbor hopping element.
While all characteristics of a certain material are determined by the
corresponding distribution $p[\epsilon_j]$, we have to keep in mind, that each 
actual sample of this material constitutes only one particular
realization $\{\epsilon_j\}$.
The disorder in the potential landscape breaks translational invariance, which
normally can be exploited for the description of ordered systems.
Hence, we have to focus on site-dependent quantities like 
the LDOS at lattice site $i$, 
\begin{equation} \label{LDOS}
  \rho_i(E) = \sum\limits_{n=1}^{N}
  | \langle i | n \rangle |^2\, \delta(E-E_n)\;,
\end{equation}
where $|n\rangle$ is an eigenstate of $H$ with energy $E_n$ and 
$|i\rangle = c^{\dag}_i |0\rangle$.
Probing different sites in the crystal, $\rho_i(E)$ will strongly fluctuate, 
and recording the values of the LDOS we may accumulate the distribution 
$f[\rho_i(E)]$.
In the thermodynamic limit, taking into account infinitely many lattice sites, 
the shape of $f[\rho_i(E)]$ will be independent of the actual 
realization of the on-site potentials $\{\epsilon_j\}$ and the chosen sites,
but depend solely on the underlying distribution $p[\epsilon_j]$.
Thus, in the sense of distributions, we have restored translational 
invariance, and the study of $f[\rho_i(E)]$ allows us to discuss the
complete properties of the Hamiltonian~\eqref{model}.
%
%Due to these fluctuations, a description of the system in terms
%of averaged values is inappropriate, and the only significant quantity is the
%distribution of the LDOS $f[\rho_i(E)]$.
%
% klingt zu negativ: macht unsere me & tydos unten zu madig
%
%
The probability density $f[\rho_i(E)]$ was found to have essentially 
different properties for extended and localized states~\cite{MF94}.
% and becomes critical at the localization transition~\cite{HT94,DPN03}.
%
For an extended state the amplitude of the wave function
is more or less uniform.
Accordingly $f[\rho_i(E)]$ is sharply peaked and symmetric about
the (arithmetically averaged) mean value $\rho_{\rm me}(E)$.
On the other hand, if states become localized,
the wave function has considerable weight only on a few sites.  
In this case the LDOS strongly fluctuates throughout the lattice and the
corresponding LDOS distribution is mainly concentrated at $\rho_i=0$,
very asymmetric and has a long tail.
The rare but large LDOS-values dominate the mean value $\rho_{\rm me}(E)$,
which therefore cannot be taken as a good approximation of the most 
probable value of the LDOS.
Such systems are referred to as `non-self-averaging'.
In numerical calculations, this different behavior may be used to 
discriminate localized from extended states in the following manner: 
Using the KPM with a variable number of moments for different interval 
sections (see Sect.~\ref{sect:KPM}) we calculate
the LDOS for a large number of samples, $K_r$, and sites, $K_s$.
Note that from a conceptual point of view, it makes
no difference if we calculate $f[\rho_i(E)]$ for one large sample and 
many sites or consider smaller (but not too small) systems and more realizations of 
disorder.
As the latter procedure is numerically more favorable, we will revert
to this.
Next, we sort the LDOS values for all energies within a small window 
around the desired energy $E$ into a histogram.
As the LDOS values vary over several orders of magnitude, a 
logarithmic partition of the histogram presents itself.
Although this procedure allows for the most intuitive determination of
the localization properties in the sense of Anderson~\cite{An58}, 
there are two drawbacks.
First, to achieve a reasonable statistics, in particular for the
slots at small LDOS values, a huge number of realizations
is necessary.
To alleviate the problems of statistical noise, it is advantageous to 
look at integral quantities like the distribution function 
\begin{equation}
F[\rho_i(E)]=\int_0^{\rho_i(E)} f[\rho^\prime_i(E)]\,\D\rho^\prime_i(E)\;,
\end{equation}
which also allows for the determination of the localization properties.
While for extended states the more or less uniform amplitudes lead to a 
steep rise of $F[\rho_i]$, for localized states the increase extends over
several orders of magnitude.
Second, for practical calculations the recording of the whole 
distribution (or distribution function) is a bit inconvenient, especially
if we want to discuss the localization properties of the whole band.
Therefore we try to capture the main features of the distribution 
by comparing two of its characteristic quantities, the 
(arithmetically averaged) mean DOS\index{mean DOS}, 
\begin{equation}
  \rho_{\text{me}}(E)  = \frac{1}{K_r K_s} \smash{
   \sum\limits_{k=1}^{K_r}\sum\limits_{i=1}^{K_s}
   \rho_i(E)} \;, \label{rhoav}
\end{equation}
and the (geometrically averaged) so called `typical' DOS \index{typical DOS},
\begin{equation}
  \rho_{\text{ty}}(E)  = \exp \left(\frac{1}{K_r K_s} \smash{
   \sum\limits_{k=1}^{K_r}\sum\limits_{i=1}^{K_s}
   \ln\bigl(\rho_i(E)\bigr)} \right) \;. \label{rhoty}
\end{equation}
The typical DOS puts sufficient weight on small values of $\rho_i$.
Therefore comparing $\rho_{\rm ty}(E)$ and $\rho_{\rm me}(E)$
allows to detect the localization transition.
We classify a state at energy $E$ with $\rho_{\rm me}(E)\neq 0$ as
localized if $\rho_{\rm ty}(E)= 0$ and as extended if $\rho_{\rm
  ty}(E)\neq 0$.
This method has been applied successfully to the pure Anderson
model~\cite{DPN03,SWWF05,ASWBF05}\index{Anderson model}
and even to more complex situations, where the effects of correlated
disorder~\cite{SWF05b}, electron-electron interaction~\cite{DK97,BHV05} 
or electron-phonon coupling~\cite{BF02,BAF04} were taken into account.

\subsection{Calculation of the Local Density of States by the
  Variable~Moment~Kernel~Polynomial~Method}
\label{sect:KPM}
\index{KPM}

\begin{figure}
  \centering 
  \includegraphics[width=0.95\linewidth,clip]{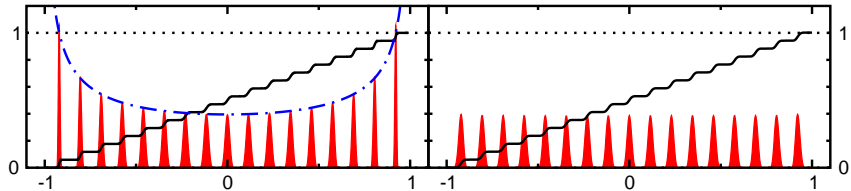}
  \caption{Approximation of a series of equally 
    weighted $\delta$-peaks using the standard KPM (left)
    and the VMKPM (right). While both methods reproduce the correct
    spectral weight (solid line), only in the latter case a uniform 
    resolution is obtained, reflected by the equal height of the peaks.
    The dashed-dotted line in the left panel is a guide to the eye 
    being proportional to the inverse of the resolution~\eqref{sigma_KPM}}
   \label{qual_VMKPM}
\end{figure}

At first glance, \eqref{LDOS} suggests that the calculation
of the LDOS\index{LDOS} could require a complete diagonalization of $H$.
It turns out, however, that an expansion of $\rho_i$ in terms of
a finite series of Chebyshev polynomials $T_n(x) = \cos(n\arccos x)$ 
allows for an incredibly precise approximation~\cite{WWAF06}.
Since the Chebyshev polynomials form an orthogonal set on $[-1,1]$,
prior to an expansion the Hamiltonian $H$ needs to be rescaled,
$\tilde{H} = (H-b)/a$. 
For reasons of numerical stability, we choose the parameters $a$ and $b$
such that the extreme eigenvalues of $\tilde H$ are $\pm0.99$.
In this way, the outer parts of the interval, where the strong 
oscillations of $T_n(x)$ can amplify numerical errors, contain no 
physical information and may be discarded.
In terms of the coefficients, the so called Chebyshev moments,
\begin{equation}
 \mu_m = \int\limits_{-1}^{1} \rho_{i}(x) T_m(x) \, \D x 
       = \sum\limits_{n=1}^{N}  \langle i  
         | n \rangle \langle n | T_m(x_n) | i \rangle 
       = \langle i | T_m({\tilde H}) | i \rangle \; , 
    \label{mu_Sp}
\end{equation}
the approximate LDOS reads
\begin{equation}
  \rho_i(x) = \frac{1}{\pi\sqrt{1-x^2}}\left( g_0\mu_0 + 
    2\sum\limits_{m=1}^{M-1} g_m\mu_m T_m(x)\right) \;.
  \label{reconstruct}
\end{equation}
The Gibbs damping factors 
\begin{equation}\label{Gibbs_Koeff}
 g_m =  \left(1-\frac{m\phi}{\pi}\right)\cos(m\phi) + \frac{\phi}{\pi}\sin(m\phi)\cot(\phi) 
\end{equation}
with $\phi = \pi/(M+1)$, are introduced to damp out the Gibbs
oscillations resulting from finite-order polynomial approximations.
The introduction of these factors corresponds to convoluting
the finite series with the so called Jackson 
kernel~\cite{SRVK96}.
In essence, a $\delta$-peak at position $x_0$ is approximated
by an almost Gaussian of width~\cite{WWAF06}
\begin{equation}
  \sigma = \sqrt{\frac{M-x_0^2(M-1)}{2(M+1)}\left(1-\cos(2\phi)\right)}  \approx 
  \frac{\pi}{M}\sqrt{1-x_0^2+\frac{4x_0^2-3}{M}}\;.
  \label{sigma_KPM}
\end{equation}
Thus for a fixed number of moments, $M$, the resolution of the expansion 
gets better towards the interval boundaries.
While in most applications this feature is harmless or even useful, here 
a uniform resolution throughout the whole band is mandatory to
discriminate resolution effects from localization.
This gets clear, if we respect that a small value of $\rho_i$ at a certain
energy may either be due to a true low amplitude of the wave function, or
to the absence of any energy level for the current disorder realization
within one kernel width.
Depending on which part of the interval we want to reconstruct, 
we need to restrict the used number of moments in (\ref{reconstruct})
accordingly to ensure a constant $\sigma$.
We call this procedure the Variable Moment KPM~(VMKPM).
The resulting approximations of a series of $\delta$-peaks
using the standard KPM and the VMKPM, respectively, are compared in 
Fig.~\ref{qual_VMKPM}.

For practical calculation of the moments, we may profit 
from the recursion relations of the Chebyshev polynomials, 
\begin{equation}
  T_{m+1}(x) = 2 x T_{m}(x) - T_{m-1}(x)\;,
\end{equation}
starting from $T_0(x)=1$ and $T_1(x)=x$, and calculate the $\mu_{m}$ iteratively.  
Additionally, we may reduce the numerical effort by another factor
$1/2$ by generating two moments with each matrix vector
multiplication by ${\tilde H}$,
\begin{equation}\label{Momcalc}
  \begin{split}
    \mu_{2m-1} & =  2
    \langle i |T_m({\tilde H})T_{m-1}({\tilde H})| i \rangle
    -  \mu_{1}\;,\\
    \mu_{2m}   & =  2
    \langle i |T_m({\tilde H})T_m({\tilde H})| i \rangle
    -  \mu_{0}  \;.
  \end{split}
\end{equation}
Note that the
algorithm requires storage only for the sparse matrix ${\tilde H}$
and two vectors of the corresponding dimension.
Having calculated the desired number of moments, we calculate several 
reconstructions~(\ref{reconstruct}) for different $M$.
We obtain the final result with uniform resolution by smoothly 
joining the corresponding results for the different subintervals.
As the calculation of the $\mu_m$ dominates the computational effort,
the additional overhead for performing several reconstructions is 
negligible as they can be done using a Fast Fourier Transform~(FFT) 
routine.

\subsection{Illustration of the Method: Anderson Localization in $3D$}
\index{Anderson localization}

\begin{figure}
  \centering 
  \includegraphics[width=0.95\linewidth,clip]{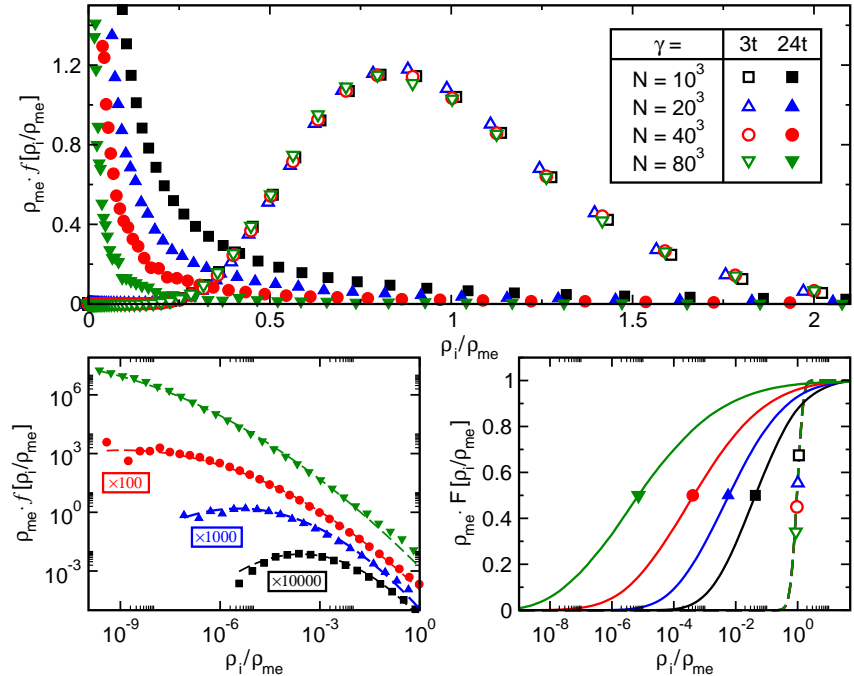}
  \caption{Upper panel: Finite-size scaling of the
    normalized LDOS distribution $f[\rho_i/\rho_{\text{me}}]$
    at the band center ($|E|\le0.01t$) for the Anderson model. 
    For the different system sizes $N$ we adapted 
    the resolution $\sigma$ such that $N\sigma\approx 6.14$ 
    is constant and calculated
    $10^6, 10^4, 1024, 400$ realizations of $\rho_i$ for 
    $N=10^3, 20^3, 40^3, 80^3$.
    Lower left panel: Double-logarithmic plot of 
    $f[\rho_i/\rho_{\text{me}}]$ 
    for the localized case together with log-normal fits 
    (\ref{log-normal}) to the data. Note 
    that for better visibility the data for $N=40^3,20^3, 10^3$
    were shifted vertically
    by $2,3,4$ orders of magnitude towards smaller values.
    Right lower panel: Distribution function 
    $F[\rho_i/\rho_{\text{me}}]$ of the
    above data}
  \label{Vdichte_Anderson}
\end{figure}

\begin{figure}[t]
  \centering 
  \includegraphics[width=0.95\linewidth,clip]{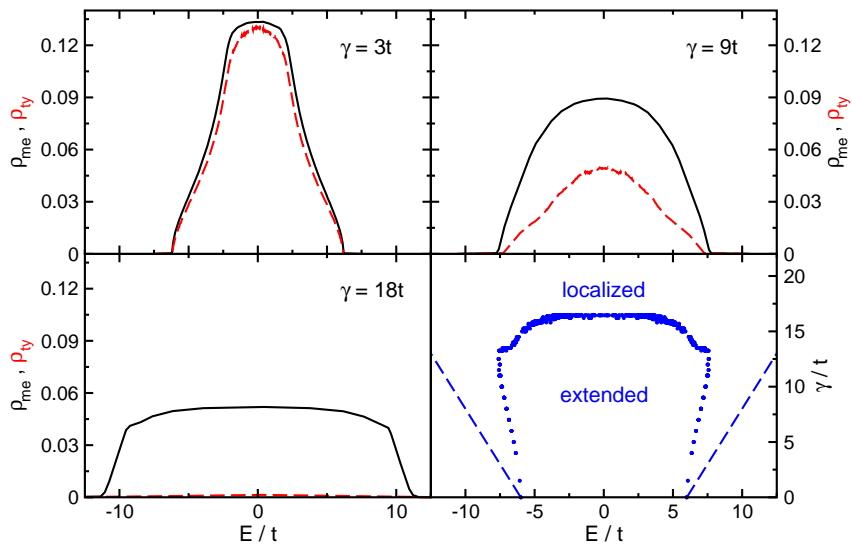}
  \caption{Mean (solid) and typical (dashed) density of states 
    of the $3D$ Anderson model on a lattice with $N=60^3$ sites
    and periodic boundary conditions~(PBC) for different values 
    of disorder $\gamma$. We used
    $N\sigma =  45$ and the shown curves result from an ensemble of
    $2000$ individual LDOS spectra obtained for different sites and
    realizations of disorder.
    Lower right panel: Mobility edge (dots) as obtained for 
    $(\rho_{\text{ty}}/\rho_{\text{me}})_c = 0.05$ and Lifshitz boundaries 
    (dashed line) }
  \label{and_3d_mety}
\end{figure} 

To clarify the power of the method, let us briefly apply it to the
Anderson model of localization \cite{An58}, for which the principal 
results are well known and can be found in the literature \cite{KM93_2,LR85}.
The Anderson model is described by the Hamiltonian~\eqref{model},  
using local potentials $\epsilon_j$, which are assumed to be independent, 
uniformly distributed random variables,
\begin{equation}\label{Kasten}
   p[\epsilon_j] = \frac{1}{\gamma} 
   \;\theta\left(\frac{\gamma}{2}-|\epsilon_j|\right)\;,
\end{equation}
where the parameter $\gamma$ is a measure for the strength of disorder.
The spectral properties of the Anderson model have
been carefully analyzed (see e.g.~\cite{FMSS85}).
% Lokalisierung fuer Unordnung & an Bandkanten
For sufficiently large disorder or near the band tails, the spectrum
consists exclusively of discrete eigenvalues, and the corresponding
eigenfunctions are exponentially localized.
% Mobilitaetskante
Since localized electrons
do not contribute to the transport of charge or energy, the energy
that separates localized and extended eigenstates is called the
mobility edge.
% 1d alles lokalisiert
For any finite amount of disorder $\gamma>0$, on a $1D$ lattice, 
all eigenstates of the Anderson Hamiltonian are localized~\cite{Bo63,MT61}.
% 2d alles lokalisiert - Skalentheorie
This is believed to hold also in $2D$, where
the existence of a transition from localized to extended states at
finite $\gamma$ would contradict the one parameter scaling
theory~\cite{AALR79,Ja98}.
% 3d beides mgl, je nach Unordnung
In three dimensions, the disorder strength has a more distinctive effect
on the spectrum. Only above a critical value $\gamma_c$ all states are 
localized, whereas for $\gamma<\gamma_c$ a pair of mobility edges 
separates the extended states in the band center from the localized ones 
near the band edges~\cite{Mo67}.
% daher 3d = unsere Wahl
For this reason, the $3D$ case serves as a prime
example on which we demonstrate how to discriminate localized from
extended states within the local distribution approach.
% wichtig: Verteilung
%Adopting Anderson's original point of view that a proper description
%of disordered systems should focus on distribution functions, we
%calculate $\rho_i(E)$ for a large number of samples, $K_r$, and
%sites, $K_s$, and study its statistical properties.
% Bildbesprechung 
In the upper panel of Fig.~\ref{Vdichte_Anderson} we show the resulting
distribution of
$\rho_i(E=0)$, normalized by its mean value $\rho_{\text{me}}$, for
two characteristic values of disorder. 
As $\rho_{\text{me}}$ is a function of disorder, this normalization 
ensures $\langle\rho_i/\rho_{\text{me}}\rangle=1$ independent of 
$\gamma$, allowing for an appropriate comparison.
% delokalisiert: symmetrisch, unempfindlich gegen Systemgroesse
In the delocalized phase, $\gamma=3t$, the distribution is rather 
symmetric and peaked close to its mean value. 
Note that increasing both the system size and VMKPM resolution, such that
the ratio of mean level spacing to the width of the Jackson kernel is fixed, 
does not change the distribution.
% lokalisiert: asymmetrisch, mean <-> probable, Systemgroesse wichtig 
This is in strong contrast to the localized phase, $\gamma=24t$, 
where the distribution of $\rho_i(E)$ is extremely asymmetric.
Although most of the weight is now concentrated close to
zero, the distribution extends to very large values of $\rho_i$,
causing the mean value to be much larger than the most probable
value.
Performing a similar finite-size scaling underlines both the
asymmetry and the singular behavior which we expect for infinite resolution
in the thermodynamic limit.
% log-normak-Verteilung
Note also, that for the localized case, the distribution of the LDOS
is extremely well approximated by a log-normal distribution~\cite{MS83},
\begin{equation}
  \Phi_{\text{log}}(x) = \frac{1}{\sqrt{2\pi\sigma_0^2}} \frac{1}{x}
  \exp\left(-\frac{\left(\ln\left(x/x_0\right)\right)^2}
    {2\sigma_0^2}\right)\;,
  \label{log-normal}
\end{equation}
as illustrated in the lower left panel of Fig.~\ref{Vdichte_Anderson}.
% Verteilungsfunktion
The shifting of the distribution towards zero for localized states is most obvious 
in the distribution function $F[\rho_i]$ which is depicted in the lower 
right panel of Fig.~\ref{Vdichte_Anderson}.
While for extended states the more or less uniform amplitudes lead to a steep rise
of $F[\rho_i]$, for localized states the increase extends over several orders of
magnitude.
% me & ty
Capturing the essential features of the LDOS distribution by concentrating on the
mean ($\rho_{\text{me}}$) and typical ($\rho_{\text{ty}}$) density of states, we
determine the localization properties for the whole energy band depending on the
disorder. 
As can be seen from Fig.~\ref{and_3d_mety},
$\rho_{\text{me}}(E)$ and $\rho_{\text{ty}}(E)$ are almost equal for
extended states, whereas for localized states $\rho_{\text{ty}}(E)$
vanishes while $\rho_{\text{me}}(E)$ remains finite.
%
% Of course, the study of entire distributions is a bit inconvenient,
% and for practical calculations, instead, we will prefer an appropriate
% statistics that uniquely characterises the distribution. The above
% findings suggest, that such a statistics is given by the arithmetic
% and geometric averages of $\rho_i(E)$.
% On the one hand, the arithmetic mean for large enough
%   $K_r$ and $K_s$ converges to the standard density of states $\rho(E)
%   = \sum_{n=1}^{N} \delta(E-E_n)$, which is not critical at the
%   Anderson transition. The geometric mean, on the other hand,
%   represents the typical value of the distribution, which, as shown
%   above, is finite in the delocalised phase, but goes to zero in the
%   localised phase.
%   
Using the well-established value $\gamma_c(E=0)\simeq 16.5t$ as a
calibration for the critical ratio $\rho_{\text{ty}}/\rho_{\text{me}}$,
required to distinguish localized from extended states 
for the used system size and resolution, we reproduce the
mobility edge in the energy-disorder plane~\cite{MK83,KM93_2,GS95}
(see lower right panel of Fig.~\ref{and_3d_mety}).
%  reentrant behavior
We also find the well-known re-entrant behavior near the unperturbed band
edges~\cite{BSK87,Qu01}: Varying $\gamma$ for some fixed values of $E$
($6t<E\le 7.6t$) a region of extended states separates two regions
of localized states.
% Lifshitz-Grenzen
The Lifshitz boundaries, shown as dashed lines,
indicate the energy range, where eigenstates are in principle allowed.
As the probability of reaching the Lifshitz boundaries is
exponentially small, we cannot expect to find states near these
boundaries for the finite ensembles considered in any numerical
calculation.

\section{Localization Effects in Quantum Percolation}
\label{sec:QPERC}
   
In disordered solids the percolation problem is characterized by the
interplay of pure classical and quantum effects.
Besides the question of finding a percolating path of `accessible'
sites through a given lattice, the quantum nature of the electrons 
imposes further restrictions on the existence of extended states and, 
consequently, of a finite DC-conductivity.

As a particularly simple model we start again from the basic 
Hamiltonian~\eqref{model} drawing the 
$\{\epsilon_j\}$ from the bimodal distribution
\begin{equation}
  p[\epsilon_j] = p\,\delta(\epsilon_j-\epsilon_A) +
  (1-p)\, \delta(\epsilon_j-\epsilon_B)\;.
  \label{binalloy}
\end{equation}
The two energies $\epsilon_A$ and $\epsilon_B$ could, for instance,
represent the potential landscape of a binary alloy A$_p$B$_{1-p}$,
where each site is occupied by an $A$ or $B$ atom with probability $p$ or
$1-p$, respectively.
Therefore we call \eqref{model} together with the distribution
\eqref{binalloy} the \emph{binary alloy model}\index{binary alloy model}.
In the limit $\Delta=(\epsilon_B-\epsilon_A)\rightarrow \infty$ 
the wave function of the $A$ sub-band vanishes
identically on the $B$-sites, making them completely inaccessible for
the quantum particles.
We then arrive at a situation where
non-interacting electrons move on a random ensemble of $\tilde{N}$
lattice points, which, depending on $p$, may span the entire lattice or not.
The corresponding \emph{quantum site percolation model}
\index{quantum site percolation model} reads
\begin{equation}\label{H_sc}
  {H} = - t \!\sum_{\langle ij \rangle \in A}\!
  ({c}_i^{\dag} {c}_j + \text{H.c.})\;,
\end{equation}  
where the summation extends over nearest-neighbor $A$-sites only and,
without loss of generality, $\epsilon_A$ is chosen to be zero.

Within the classical percolation scenario the percolation threshold $p_c$
is defined by the occurrence of an infinite cluster $A_\infty$ of
connected $A$ sites.
For the simple cubic lattice this site-percolation
threshold \index{site percolation threshold} 
is $p_c^{3D}=0.311609$~\cite{BFMSPR99} for the $3D$ case and
$p_c^{2D}=0.592746$~\cite{NZ00} in $2D$.
In the quantum case, the
multiple scattering of the particles at the irregular boundaries of
the cluster can suppress the wave function, in particular within
narrow channels or close to dead ends of the cluster.
Hence, this
type of disorder can lead to absence of diffusion due to localization,
even if there is a classical percolating path through the crystal.
On
the other hand, for finite $\Delta$ the tunneling between $A$ and $B$
sites may cause a finite DC-conductivity although the $A$ sites are not
percolating.
Naturally, the question arises whether the quantum
percolation threshold $p_q$, given by the probability above which an
extended wave function exists within the $A$ sub-band, is larger or
smaller than $p_c$.
Previous results~\cite{SLG92,NBR02} for finite values
of $\Delta$ indicate that the tunneling effect has a marginal
influence on the percolation threshold as soon as $\Delta\gg 4tD$.

\subsection{$3D$ Site Percolation}
\label{sec:perc_3d}

\begin{figure}%[ht]
  \centering 
  \includegraphics[width=0.95\linewidth]{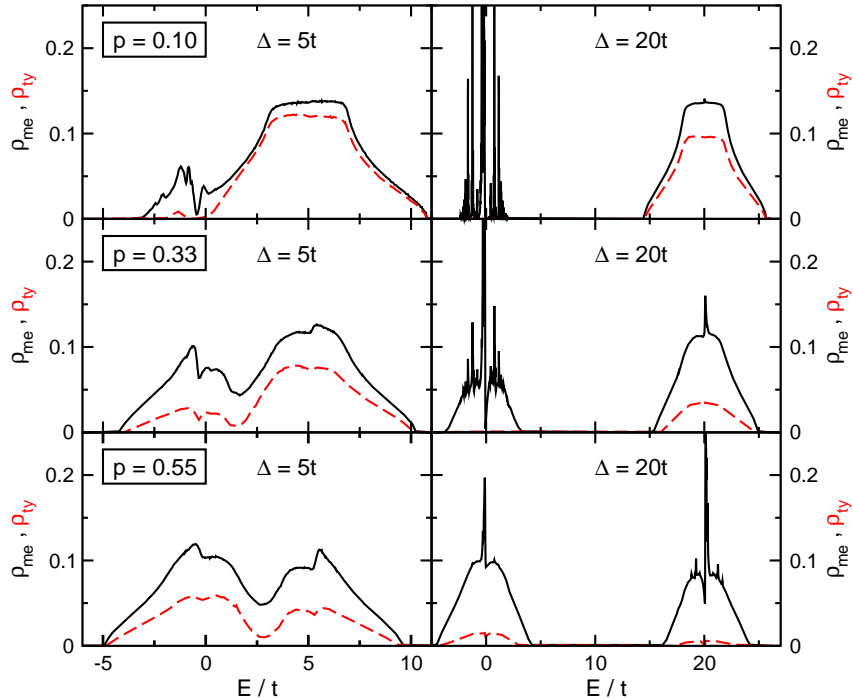}
   \caption{Mean (upper solid line) and typical (lower dashed line) 
     DOS of the binary alloy model on a $N=64^3$ lattice with PBC for different 
     concentrations of $A$-sites
     $p$ and local potential differences $\Delta$. Taking all sites into 
     account, we chose the resolution such that 
     $N\sigma=45$ and calculated $1000$ individual LDOS spectra for different
     probe realizations and sites}\label{endl_delta}
\end{figure}    

\begin{figure}%[ht]
  \centering 
  \includegraphics[width=0.95\linewidth]{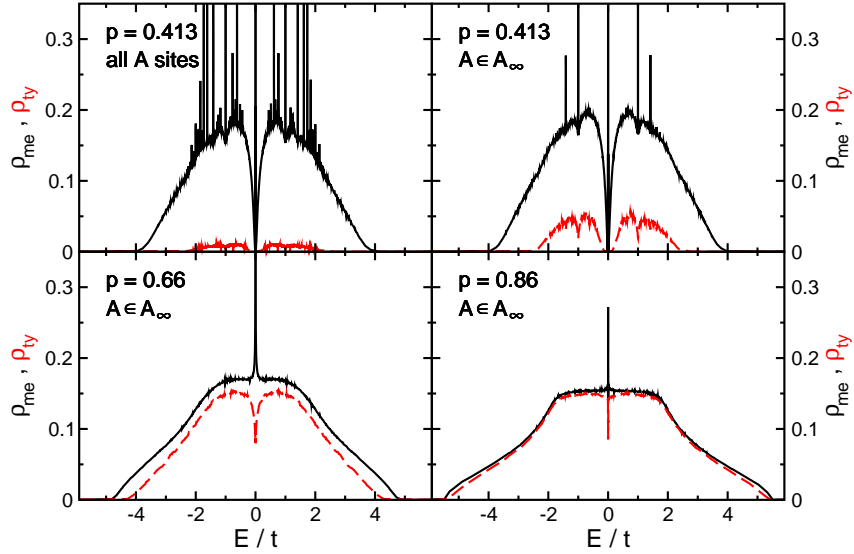}
  \caption{Mean (solid line) and typical (dashed line)
    DOS for the quantum percolation model in the limit
    $\Delta\to\infty$.  While in the upper left panel all $A$-sites
    are taken into account, the other three panels show data for the
    restricted model on the spanning cluster $A_\infty$
    only (note that  $\rho_{\rm ty}$ is smaller in the former case
    because there are more sites with vanishing amplitude 
    of the wave function). Using $\sigma=0.0002$, we adapt the system
    sizes to ensure $\bar{N}\sigma=45$ for the $A_\infty$ case, i.e. , 
    $N=64^3, 70^3, 84^3$ for $p=0.86, 0.66, 0.413$. For 
    the case of all $A$ sites at $p=0.413$,  $N=82^3$
    guarantees $\tilde{N}\sigma=45$.
    For each system, we consider $512$ individual LDOS spectra
    for different sites and realizations of disorder
  }\label{perc_3d_mety}
\end{figure}

Before discussing possible localization phenomena let us investigate
the behavior of the mean DOS for the binary alloy and 
quantum percolation model in $3D$.
Figure~\ref{endl_delta} shows
that as long as $\epsilon_A$ and $\epsilon_B$ do not differ too much
there exists a single and if $p\neq 0.5$ asymmetric
electronic band~\cite{SLG92}.
At about $\Delta \simeq 4tD$ this band
separates into two sub-bands centered at $\epsilon_A$ and $\epsilon_B$,
respectively.
The most prominent feature in the split-band regime is
the series of spikes at discrete energies within the band.
As an
obvious guess, we might attribute these spikes to eigenstates on
islands of $A$ or $B$ sites being isolated from the main
cluster~\cite{KE72,BA96}.
It turns out, however, that some of the
spikes persist, even if we neglect all finite clusters and restrict
the calculation to the $\bar{N}$ sites of $A$-type on the spanning 
cluster, $A_\infty$.
This is illustrated in the upper panels of Fig.~\ref{perc_3d_mety}, where we
compare the DOS of the binary alloy model at $\Delta\to\infty$
and the quantum percolation model. 
Increasing the concentration of accessible sites
the mean DOS of the spanning cluster is evocative of the DOS of the
simple cubic lattice, but even at large values of $p$ a sharp peak
structure remains at $E=0$ (see Fig.~\ref{perc_3d_mety}, lower panels).

\begin{figure}%[ht]
  \centering 
  \includegraphics[width=0.95\linewidth]{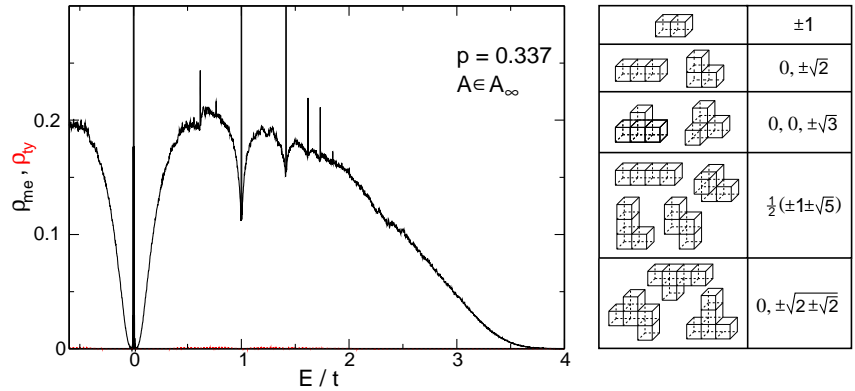}
  \caption{Left: Mean (solid line) and typical (dashed line)
    DOS for the quantum percolation model with $p=0.337$ on a $N=100^3$ lattice
    (restriction to $A_\infty$, PBC, $\bar{N}\sigma=45$). The data
    are an ensemble average over 1000 realizations of disorder and different sites.
    Note that $\rho_{\text{ty}} <10^{-3}$ holds in the whole band.  Right:
    Some cluster configurations related to the special energies at
    which the peaks in $\rho_{\text{me}}$ occur}\label{Cl_Tab}
\end{figure}

To elucidate this effect, which for a long time was partially not 
accounted for in the literature~\cite{KE72,SLG92,OOM80,SEG87}, 
in more detail, in Fig.~\ref{Cl_Tab} we fix $p$ at $0.337$, shortly
above the classical percolation threshold.
In addition to the most dominant peaks at $E/t = 0, \pm 1,
\pm\sqrt{2}$, we can resolve distinct spikes at $E/t =
\frac{1}{2}\left(\pm 1\pm \sqrt{5}\right), \pm \sqrt{3}, \pm\sqrt{ 2\pm 
\sqrt{2}}, \ldots$ .
These special energies coincide with the
eigenvalues of the tight-binding model on small clusters of the
geometries shown in the right part of Fig.~\ref{Cl_Tab}. 
In accordance with~\cite{KE72} and~\cite{CCFST86} we can thus
argue that the wave functions, which correspond to these special
energies, are localized on some `dead ends' of the spanning cluster.

The assumption that the distinct peaks correspond to localized
wave functions is corroborated by the fact that the typical DOS
vanishes or, at least, shows a dip at these energies.
Occurring also
for finite $\Delta$ (Fig.~\ref{endl_delta}), this effect becomes more
pronounced as $\Delta\to\infty$ and in the vicinity of the classical
percolation threshold $p_c$.
From the study of the Anderson model we know that localization leads at
first to a narrowing of the energy window containing extended states.
For the percolation problem, in contrast, with decreasing $p$ the typical
DOS indicates both localization from the band edges and localization at 
particular energies within the band.
Since finite cluster wave functions like
those shown in Fig.~\ref{Cl_Tab} can be constructed for numerous
other, less probable geometries~\cite{Sc03}, an infinite discrete series
of such spikes might exist within the spectrum, as proposed in~\cite{CCFST86}.
The picture of localization in the quantum percolation model
is then quite remarkable.
If we generalize our numerical data for the
peaks at $E=0$ and $E =\pm t$, it seems as if there is an infinite
discrete set of energies with localized wave functions, which is dense
within the entire spectrum. 
In between there are many extended states, but to avoid mixing, 
their density goes to zero close to the localized states.
Facilitated by its large special weight (up to 11\% close to $p_c$) this is
clearly observable for the peak at
$E=0$, and we suspect similar behavior at $E =\pm t$.
For the other
discrete spikes the resolution of our numerical data is still too poor
and the system size might be even too small to draw a definite conclusion.

\begin{figure}
  \centering 
  \includegraphics[width=0.95\linewidth]{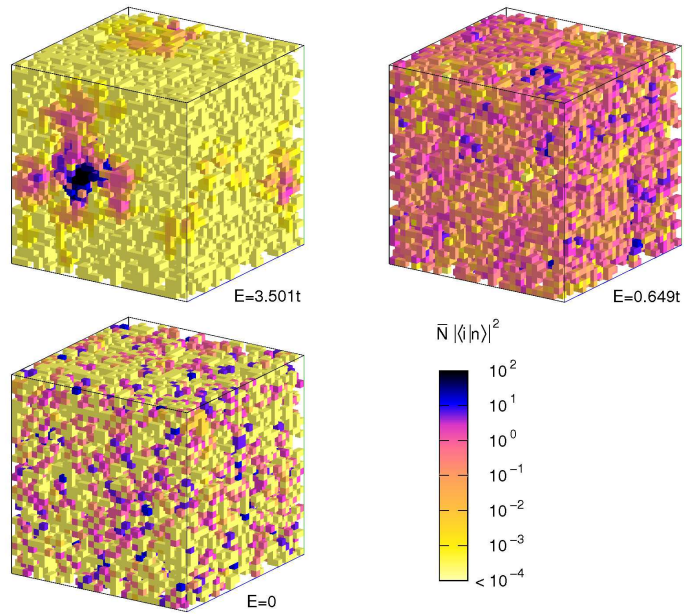}
  \caption{Normalized probability density $\bar{N}|\langle i|n\rangle|^2$ 
    for three eigenstates of the quantum percolation model.
    The chosen energies are representative for a localized ($E=3.501t$), 
    extended ($E=0.649t$) and anomalous localized ($E=0$) state.
    We performed an exact diagonalization for $A_\infty$ for one 
    realization of disorder on a $N=32^3$ lattice with PBC and $p=0.458$} 
    \label{CharZst}
\end{figure}

In order to understand the internal structure of the extended and
localized states we calculate the probability density of 
specific eigenstates of~\eqref{H_sc} restricted to $A_\infty$ for a
fixed realization of disorder. 
Figure~\ref{CharZst} visualizes the spatial variation of 
$|\langle i|n\rangle|^2$ for an occupation probability well above the 
classical percolation threshold. 
The figure clearly indicates that the state with $E =0.649t$
is extended, i.e. the spanning cluster is quantum mechanically
`transparent'. 
On the contrary, at $E=3.501t$, the wave function is
completely localized on a finite region of the spanning cluster. 
This is caused by the scattering of the particle at the random surface 
of the spanning cluster.

A particularly interesting behavior is observed at $E=0$. 
Here, the eigenstates are highly degenerate and we can form wave functions
that span the entire lattice in a checkerboard structure with zero and 
non-zero amplitudes (see Fig.~\ref{CharZst}). 
Although these states are extended in the sense that they are not confined 
to some region of the cluster, they are localized in the sense that they 
do not contribute to the DC-conductivity. 
This is caused by the alternating structure which suppresses the 
nearest-neighbor hopping, and in spite of the high
degeneracy, the current matrix element between different $E=0$ states
is zero.
Hence, having properties of both classes of states these
states are called anomalously localized~\cite{SAH82,ITA94}.
%
%The checkerboard structure is also observed in $2D$,
%but is less pronounced than in the $3D$ case,
%due to the reduced spectral weight of the central peak.
%
Another
indication for the robustness of this feature is its persistence for
mismatching boundary conditions, e.g., periodic (anti-periodic)
boundary conditions for odd (even) values of the linear extension $L$. 
In these cases the checkerboard is matched to itself by a
manifold of sites with vanishing amplitude.

Furthermore for the previously mentioned special energies, 
the wave function vanishes identically except for some finite domains on 
loose ends (like those shown in the right panel of Fig.~\ref{Cl_Tab}),
where it takes, except for normalisation, the values ($\pm 1, \mp 1$), 
($1, \pm\sqrt{2},1$), ($-1,0,1$),~$\ldots$
Note that these regions are part of the spanning cluster, connected to the
rest of $A_\infty$ by sites with wave function amplitude zero~\cite{CCFST86}.
\begin{figure}
  \centering 
  \includegraphics[width=0.95\linewidth,clip]{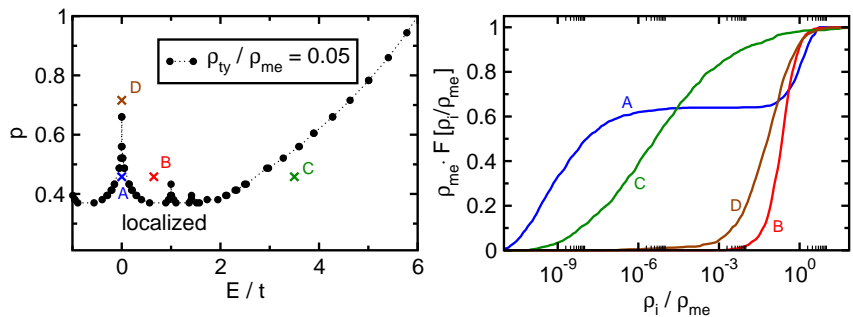}
  \caption{Left panel: Normalized typical DOS 
    $\rho_{\rm ty}/\rho_{\rm me}$ in the concentration-energy plane for the 
    quantum percolation model restricted to $A_\infty$. For a fixed resolution of
    $\sigma=0.0002$ we adapted the system sizes to keep $\bar{N}\sigma=45$ 
    constant, i.e., $N=62^3,\ldots,120^3$ for $p=0.944,\ldots,0.318$.
    The results base on the calculation of 1000 individual LDOS spectra 
    for different cluster realizations and sites.
    Right panel: Distribution function of the LDOS for four characteristic
    states, for which the parameters $(p,E)$ are indicated by crosses 
    in the left panel. The parameters for the curves $A,B,C$ are the same
    for which we showed the characteristic states in Fig.~\ref{CharZst}
    \label{perc_3d_mobedge}
    }
\end{figure}

In the past most of the methods used in numerical studies of Anderson
localization have also been applied to the binary alloy model and the
quantum percolation model
in order to determine the quantum percolation threshold $p_q$, 
defined as the probability $p$ below which all states are localized 
(see, e.g.,~\cite{MDS95,KN90} and references therein).
The existence of $p_q$ is still disputed.
As yet the results for $p_q$ are far less precise than, e.g., 
the values of the critical disorder reported for the Anderson model.
For the simple
cubic lattice numerical estimates of quantum site-percolation
thresholds range from 0.4 to 0.5 (see~\cite{KN90} and references 
therein).
In Figs.~\ref{endl_delta}-\ref{Cl_Tab} we present data for 
$\rho_{\rm  ty}$ which shows that $p_q>p_c$.
In fact, within numerical accuracy,
we found $\rho_{\rm ty}=0$ for $p=0.337>p_c$.

To get a more detailed picture we calculate the normalized typical
DOS, $\rho_{\rm ty}/\rho_{\rm me}$, in the whole
concentration-energy plane.
The left panel of Fig.~\ref{perc_3d_mobedge} presents such kind
of phase diagram of the quantum percolation model.
This data also supports a finite quantum percolation threshold 
$p_q\stackrel{>}{\sim} 0.4>p_c$ (see also~\cite{SLG92,KN90,KSHS97,KO99}), but as
the discussion above indicated, for $E=0$ and $E=\pm t$ the critical
value $p_q(E) = 1$, and the same may hold for the set of other
`special' energies.
The transition line between localized and extended states, $p_q(E)$, might 
thus be a rather irregular (fractal?) function.
On the basis of our LDOS distribution approach, however,
we are not in the position to answer this question with full rigor. 

% \begin{figure}]
%   \centering 
%   \includegraphics[width=0.95\linewidth]{fig7.eps}
%   \caption{Characteristic probability distributions 
%     of the LDOS.}\label{DistFkt}
% \end{figure}

Finally let us come back to the characterization of extended and
localized states in terms of distribution functions.
The right panel of Fig.~\ref{perc_3d_mobedge} displays the distribution function,
$F[\rho_i(E)]$, for four typical points in the energy-concentration plane, 
corresponding to localized, extended, and anomalously localized states, respectively.
The differences in $F[\rho_i(E)]$ are significant.
The slow increase of $F[\rho_i(E)]$ observed for localized states
corresponds to an extremely broad LDOS-distribution, 
with a very small most probable (or typical) value of $\rho_i(E)$.
This is in agreement with the findings for the Anderson model.
Accordingly the jump-like increase found for extended states is
related to an extremely narrow distribution of the LDOS centered around
the mean DOS, where $\rho_{\rm me}$ coincides with the most probable
value.
At $E=0$ and low $p$, the distribution function exhibits two steps,
leading to a bimodal distribution density. Here the first (second)
maximum is related to sites with a small (large) amplitude of the
wave function -- a feature that substantiates the checkerboard 
structure discussed above.
For higher $p$, where we already found a reduced spectral weight of the
central peak in $\rho_{\text{me}}$, also the two step shape of the distribution
function is less pronounced.
Therefore we may argue, the increase in weight of the central peak for lower $p$
is substantially due to the checkerboard states.

Having a rather complete perception of the physics in $3D$, 
let us now come to the $2D$ case, for which the findings in
literature are more controversial.

\subsection{$2D$ Site Percolation}
\label{sec:perc_2d}

Although the main characteristics of the $3D$ site-percolation
problem, e.g., the fragmentation of the spectrum, persist
in $2D$, there exist some particularities and additional difficulties. 
In particular, the existence of a quantum percolation threshold
$1>p_q>p_c=0.592746$ is even less settled than in $3D$.
Published estimates range from $p_q=p_c$ to $p_q=1$ (see~\cite{KN90,IN07p}
and references therein). 
This uncertainty is due to the extremely large
length scales, on which localization phenomena take place in $2D$,
a fact well-known for the standard Anderson model. 
Furthermore the special characteristics of the band center 
states seem to be of particular importance~\cite{ITA94,ERS98b}.
\subsubsection{Local Density of States}
\begin{figure}[b]
  \centering 
  \includegraphics[width=0.95\linewidth]{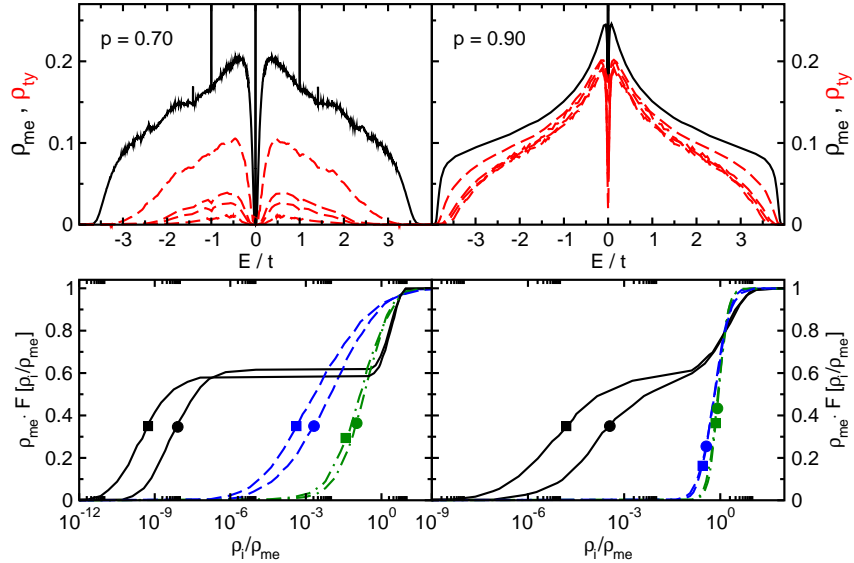}
  \caption{Upper panels: Mean (solid line) and typical (dashed lines)
    DOS for the $2D$ percolation model for
    $p=0.7$ (left) and $p=0.9$ (right).
    The dashed lines correspond to the typical DOS for different
    system sizes (from top to bottom) $N=114^2, 400^2, 572^2, 1144^2$ for 
    $p=0.7$ and $N=100^2, 350^2, 500^2, 1000^2$ for $p=0.9$. Again,
    we adapt $\sigma$, such that $\bar{N}\sigma = 45$.
    The curves are ensemble averages over
    $1000$ ($500$ for the largest systems) individual LDOS spectra for
    different cluster realizations and sites.
    Lower panel: Distribution function of the LDOS for three characteristic
    energies: $E=0$ (solid lines), $E=0.5t$ (dashed-dotted lines), and $E=2.5t$
    (dashed lines). To underline the finite size scaling, we show
    the results for two system sizes, $350^2$ ($400^2$) marked with 
    circles, and  $500^2$ ($572^2$) with squares
}\label{perc_2d_mety}
\end{figure}
Figure~\ref{perc_2d_mety} shows the mean and typical DOS of the
$2D$ quantum percolation model, calculated by the LD approach.
For large $p$, $\rho_{\text{me}}$ clearly resembles the
$2D$ DOS shape for the ordered case.
For these parameters $\rho_{\text{ty}}$ nearly coincides with
$\rho_{\text{me}}$ except for the band center, where $\rho_{\text{ty}}$ 
shows a strong dip.
If we reduce the occupation probability, the spikes at special energies appear
again (see Sect.~\ref{sec:perc_3d}), with  most spectral weight at $E=0$.
The weight of the central peak
(9\% close to $p_c$) is reduced as compared to the $3D$ case.

In order to obtain reliable results for the infinite system,
we examine the dependency of $\rho_{\text{ty}}$ on the system size,
for fixed $\bar{N}\sigma$. 
Here, we find a characteristic difference between large and small $p$.
Whereas for large $p$, above a certain system size, $\rho_{\text{ty}}$ is 
independent of $N$, it continuously decreases for low occupation
probabilities.
This behavior is evocative of extended and localized states, respectively.
%
%%%%%%%%%%%%%%%%
%
Taking a look at the underlying distribution functions, we find a similar
situation as in the $3D$ case.
At $E=0$, the two level distribution evolves, indicating the checkerboard 
structure of the state.
Away from the special energies, the distribution function exhibits the 
shape characteristic for extended and localized states.
This behavior exposes when we compare $F[\rho_i(E)]$ for different system
sizes.
Whereas for extended states the distribution function is insensitive against
this scaling, it shifts towards smaller values for localized ones.

Although these results are quite encouraging, one aspect deserves further
attention.
If we try to calculate the LDOS distribution at a given energy $E$, due to the
finite resolution of the KPM, it will also contain contributions from states
in the vicinity of $E$.
Thus taking the fragmentation of the spectrum \index{fragmentation of the spectrum}
into localized and extended states seriously, the LDOS distribution within this 
artificial interval will contain contributions of each class of states. 

For practical calculations, this causes no problems, as except for the most 
pronounced peaks, the probability of finding a state which is localized on
one of the geometries like in the right panel of Fig.~\ref{Cl_Tab} drops very 
fast with its complexity.

%A method which circumvents this conceptual problem, 
%requires in principle an arbitrary high energy resolution.
%
%To this extend, in the next subsection, we propose another approach 
%to the $2D$ quantum percolation problem, which is also based on a Chebyshev 
%expansion.
%
\subsubsection{Time Evolution}
\index{time evolution}
The expansion of the time evolution operator $U(\tau) = \E^{-\imag H\tau}$ in 
Chebyshev polynomials allows for a very efficient method to calculate
the dynamics of a quantum system.
We may profit from this fact by calculating the recurrence probability of
a particle to a given site, $P_R(\tau)$, which for $\tau\to\infty$ may serve as
a criterion for localization~\cite{Ja98,KM93_2}.
While for extended states on the spanning cluster $P_R(\tau\to\infty)=1/\bar{N}$,
which scales to zero in the thermodynamic limit, a localized state will have
a finite value of $P_R(\tau)$ as $\bar{N}\to\infty$.
%
%The advantage of considering the time evolution is that we may fix the
%energy of our initial state to arbitrary precision.
%
%In general this initial state is not an eigenstate of the system and therefore
%contains contributions of the whole spectrum.
The advantage of considering the time evolution is that 
in general the initial state is not an eigenstate of the system and therefore
contains contributions of the whole spectrum.
This allows for a global search for extended states and a detection of $p_q$. 

% and, provided
%we have a suitable algorithm for the time evolution, the inevitable 
%energy shift due to numerical time propagation may be controlled by the
%used time step.

%
Let us briefly outline how to calculate the time evolution of the system
by means of Chebyshev expansion~\cite{TK84,CG99}\index{Chebyshev expansion}.
%
% Starting from the time dependent Schr\"odinger equation (for convenience we set $\hbar=1$)
% \begin{equation}
%   \imag \frac{\partial}{\partial\tau}|\psi(\tau)\rangle = H |\psi(\tau)\rangle\;,
% \end{equation}
% we may propagate an initial state $|\psi_0\rangle$ in time by 
% applying the time evolution operator $U(\tau) = \E^{-\imag H\tau}$.
% %
% If we had exact knowledge of $U(\tau)$, we could calculate the resulting wave function 
% for arbitrary times at once.
% %
% However, the calculation of the exponential of an operator requires its complete
% diagonalization, and this is not possible for large systems.
% %
% Thus, the standard procedure is to approximate $U(\tau)$ in some way and to cover
% the desired time interval by several small time steps $\Delta\tau$, for which the 
% time evolution is still well described by the approximate $U(\tau)$.
% %
% The probably most common of these approximations is the
% Crank-Nicolson scheme~\cite{PTVF92}, in which the wave function at the new time step
% is calculated as
% \begin{equation}
%  \left(1+ \tfrac{\imag}{2} H \Delta\tau\right)|\psi(\tau+\Delta\tau\rangle = 
%  \left(1- \tfrac{\imag}{2} H \Delta\tau\right)|\psi(\tau)\rangle\;.
% \end{equation}
% %
% Note, that besides the matrix vector multiplication (MVM) on the right hand side, in each 
% time step also the solution of a (sparse) system of linear equations is required to get 
% $|\psi(\tau+\Delta\tau\rangle$.
%
%In contrast to this, for the expansion of $U(\tau)$ in terms of Chebyshev polynomials only
%MVM are required, being a big advantage in terms of computing time.
%
Of course, as a first step, the Hamiltonian has to be rescaled to the definition interval
of the Chebyshev polynomials, leading to
\begin{equation}
  U(\tau) =  \E^{-\imag (a\tilde{H} + b) \tau} = \E^{-\imag b \tau} 
  \left[ c_0 + 2\sum\limits_{k=1}^{M} c_k T_k(\tilde{H}) \right]\;.
\end{equation}
The expansion coefficients $c_k$ are given by 
\begin{equation}
  c_k = \int\limits_{-1}^1 \dfrac{T_k(x)e^{-\imag ax\tau}}{\pi \sqrt{1-x^2}}\D x = (-\imag)^k J_k(a\tau)\;,
\end{equation}
where $J_k$ denotes the Bessel function of the first kind.
Due to the fast asymptotic decay of the Bessel functions
\begin{equation}
  J_k(a\tau) \sim \frac{1}{k!} \left(\frac{a\tau}{2}\right)^k \sim
  \frac{1}{\sqrt{2\pi k}} \left( \frac{\E a\tau}{2k} \right)^k \quad \text{for}\quad k\to \infty\;,
\end{equation}
the higher-order expansion terms vanish rapidly.
Thus we do not need additional damping 
factors as in the Chebyshev expansion of the LDOS, but may truncate the series at 
some order without having to expect Gibbs oscillations.
Note that the expansion order necessary to obtain precise results will surely depend on 
the time step used, normally $M\sim 2at$ will be sufficient.
Thus, for optimum performance of the algorithm, we have to find a 
suitable compromise between time step $\Delta\tau$  and expansion 
order $M$.
Anyhow, for reasonable $M$, the method permits the use of larger time steps compared to the 
standard Crank-Nicolson algorithm.

\begin{figure}
  \centering 
  \includegraphics[width=0.95\linewidth]{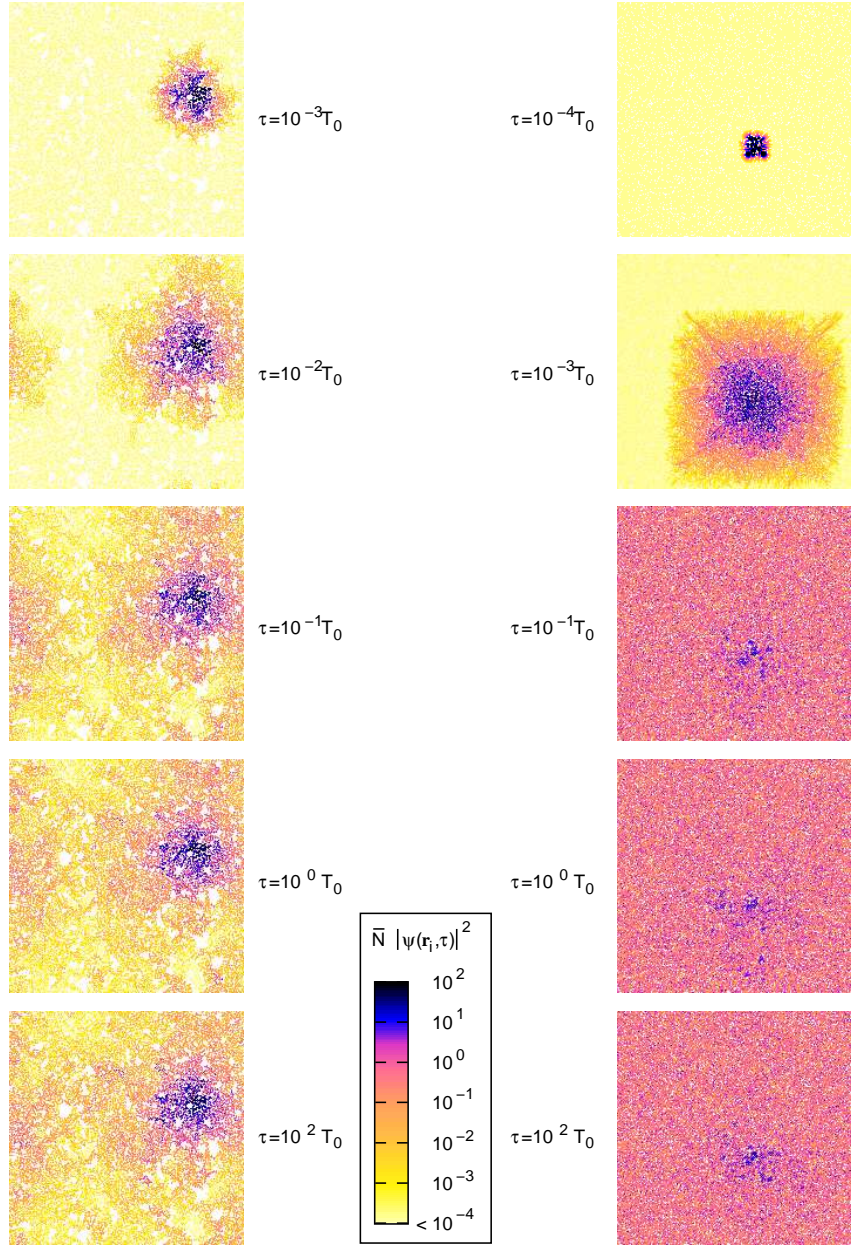}
  \caption{Time evolution of a state at $E=0.5t$ on the spanning cluster of
    a $N=200^2$ lattice for $p=0.65$ (left) and $p=0.90$ (right).  We show
    the normalized square of the wave function amplitude at the different sites 
    $\bar{N} |\psi({\mathbf r}_i,\tau)|^2$ . Due to this normalization, for an extended
    state which is evenly spread over all sites of $A_\infty$ this
    quantity is equal to unity}
  \label{perc_2d_waveampl}
\end{figure}

Having this powerful tool at hand, we are now in the position to calculate 
how $\psi({\mathbf r}_i,\tau)$ evolves on $A_\infty$ in time.
The 'natural' time scales of the system are given by the energy of the 
nearest-neighbor hopping element $\tau_0 = 1/t$, describing one hopping event, and 
the time a particle needs (in a completely ordered system) to visit each site 
once, $T_0=N\tau_0$.
As initial state, we prepare a completely localized state, whose amplitude
vanishes exactly, except for two neighboring sites, where it has amplitudes $a$ and 
$\sqrt{1-a^2}$, respectively.
For this state we can calculate the energy $E = 2ta\sqrt{1-a^2}$ and, 
choosing $a$ appropriately, we may continuously tune $E\in[-t,t]$.
%
%Clearly for the two-site cluster we cannot reach the ground state energy ($-2t$)
%of a tight binding electron in the $1D$ ordered case.
%
Taking into account more complicated initial configurations of occupied sites 
(see right panel in Fig.~\ref{Cl_Tab}) we may also adjust higher energies.
For each starting position, however, the local structure of $A_\infty$ 
limits the possible configurations.

In Fig.~\ref{perc_2d_waveampl} we compare the time evolution 
of a state for high and low occupation probability $p$,
for which qualitatively different behaviors emerge. 
For $p=0.65$, the wave function is localized on a finite region of the cluster.
Following the time evolution up to very long times ($>100T_0$)
we demonstrate that this is not just a transitional state during the
spreading process of the wave function, but true 'dynamical' localization.
This behavior is in strong contrast to $p=0.90$, where the state spreads over 
the whole cluster within a short fraction of this time ($0.1T_0$).
%
%For longer times, the fluctuations in the amplitude are damped out and
%in the limit $\tau\to\infty$ all sites will be reached with equal probability.
%
%For any fixed times, we still see some sites with slightly larger 
%amplitudes (see. the darker dots in the last image of the time series).
% 
%Their positions, however, are not fixed but fluctuate from one time step to the next.
%
%This behavior is consistent with the notion of an extended state, and a short 
%time average shows a completely uniform wave function.
%
For any fixed time there are some sites with slightly larger
amplitudes (the darker dots in the last image of the time series).
Those are due to contributions from localized states which are also present in the
initial state.
However, as the wave function extends over the whole cluster, for large $p$
not all states in the spectum may be localized.
Since the time evolution of initial states at $p=0.65$ and $p=0.90$ behaves in 
such a different manner, we conclude that there exists a quantum percolation 
threshold in between.

Consequently, we have a method at hand, capable to visualize the dynamical 
properties of localization in quantum percolation~\cite{SF08}.

\section{Percolative Effects in Advanced Materials}
\label{sec:adv_mat}
Current applications of quantum percolation concern e.g. 
transport properties of doped semiconductors~\cite{ITA94} and 
granular metals~\cite{FIS04}, metal-insulator transition 
in 2D n-GaAs heterostructures~\cite{SLHPWR05}, 
wave propagation through binary inhomogeneous media~\cite{AL92},
superconductor-insulator and (integer) quantum Hall
transitions~\cite{DMA04,SMK03}, or the dynamics of atomic Fermi-Bose
mixtures~\cite{SKSZL04}. Another important example is the
metal-insulator transition in perovskite manganite films and the related
colossal magnetoresistance (CMR) effect,\index{CMR effect} which 
in the meantime is believed
to be inherently percolative~\cite{BSLMDSS02}. Quite recently
(quantum) percolation models have been proposed to
mimic the minimal conductivity in undoped graphene~\cite{CFAA07p}\index{graphene}. 
In doped graphene, in the split-band regime,  an internal mobility 
edge might appear around the Fermi energy by introducing impurities~\cite{Na07p}.
Moreover, geometric disorder is shown to strongly affect electronic transport 
in graphene nanoribbons~\cite{MB07p}.  
In the remaining part of this paper we exemplarily investigate two 
specific random resistor network models to describe qualitatively charge 
transport in graphene sheets and bulk manganites.

\subsection{Graphene}
\index{graphene}

Due to its remarkable electronic properties, recently a lot of activity has been 
devoted to graphene, the atomic mono/bi-layer modification of 
graphite~\cite{NGMJZDGF04,NMMFKZJSG06,GN07}.
Especially the gapless spectrum with linear dispersion
near the Fermi energy and the possibility of continuously varying the charge
carrier density (from n- to p- type) by means of applying a gate voltage
are of technological interest. 
In view of possible applications, like graphene-based field effect
transistors, it is highly desirable to know how these
characteristics change in the presence of disorder, inherent in any prepared probe.
Therefore much work has been dedicated to study possible localization effects due
to the presence of disorder (cf.~\cite{PGLPC06} and references therein).

The extraordinary electronic structure of graphene results in unusual transport properties.
In this material a finite minimal conductivity is observed, which might be attributed to 
a mesoscopically inhomogeneous density of charge carriers~\cite{Ge06,MAULSVY07p,CF07p},
caused by spatially varying charge trapping on the substrate.
To describe the influence of these charge inhomogeneities on the transport 
properties, percolative random resistor networks~(RRN)\index{random resistor network}
\index{RRN} have been proposed~\cite{CFAA07p}.
Following this line, we apply the LD approach to a minimal model~\cite{SF08b}
that can be constructed in generalization of the $2D$ percolation model described
in Sect.~\ref{sec:perc_2d}.

\begin{figure}[b]
  \centering 
  \includegraphics[width=0.95\linewidth]{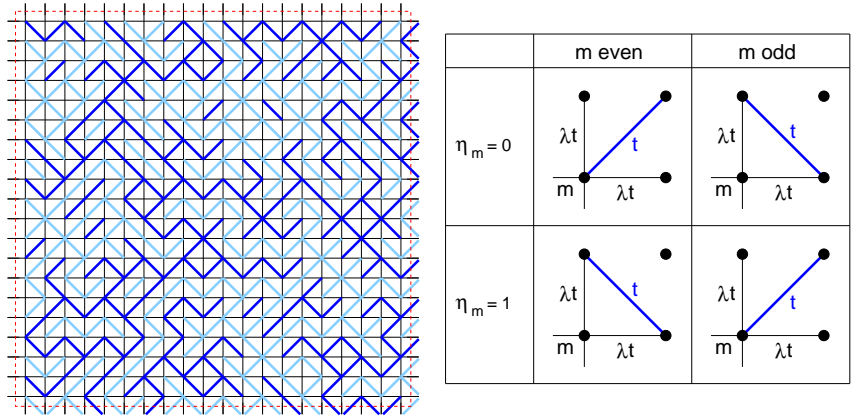}
  \caption{Left panel: One particular cluster realization of the 
    RRN model at $p=0.5$. Right panel: Visualization of the generation rule 
    for the RRN model}
  \label{fig:RRN_model}
\end{figure}

Let us consider a $2D$ lattice on which the sublattices 
represent regions of different charge carrier concentrations.
These regions (sites) are randomly connected with each other
(left panel of Fig.~\ref{fig:RRN_model}).
The hopping probability for such links (to some next nearest neighbors) 
is assumed to be much higher than for direct nearest neighbor hopping events.
The later ones are reduced by the leakage, $\lambda$, as compared to the others.
To examine the influence of anisotropic hopping, we model the corresponding
RRN by the Hamiltonian
\begin{equation}
  \begin{split}
    H  =  -t\Big[ \lambda  \sum \limits_{m} \left( c^{\dag}_m c_n +  
        c^{\dag}_m c_e \right) +  \sum \limits_{m\;\text{even}} \left(\eta_m c^{\dag}_m c_{ne} +  
        (1-\eta_m) c^{\dag}_e c_n\right) + \\
      +  \sum \limits_{m\;\text{odd}}  \left( \eta_m c^{\dag}_e c_n + 
        (1-\eta_m) c^{\dag}_m c_{ne} \right)\Big] + \text{H.c.}\;,
  \end{split}
\label{model_RRN}
\end{equation}
where $e$(ast) and $n$(orth) denote the nearest neighbors of site $m$ in $x$- and
$y$-direction and $ne$ is the next-nearest neighbor in `north-east' direction.
We assume PBC.
The random variable $\eta_m\in\{0,1\}$ determines which diagonal in a square is 
connected (right panel in Fig.~\ref{fig:RRN_model}).
Shifting the expectation value of the $\{\eta_m\}$, we may adjust the 
anisotropy of the system $p=\langle\eta_m\rangle$. 
While the link directions are isotropically distributed for $p=0.5$, 
increasing (decreasing) $p$ generates a preferred direction of hopping, 
favoring stripe-like structures.
Note that in the limit of vanishing leakage this model is equivalent to 
the $2D$ percolation model discussed in Sect.~\ref{sec:QPERC}.

\begin{figure}
  \centering 
  \includegraphics[width=0.95\linewidth]{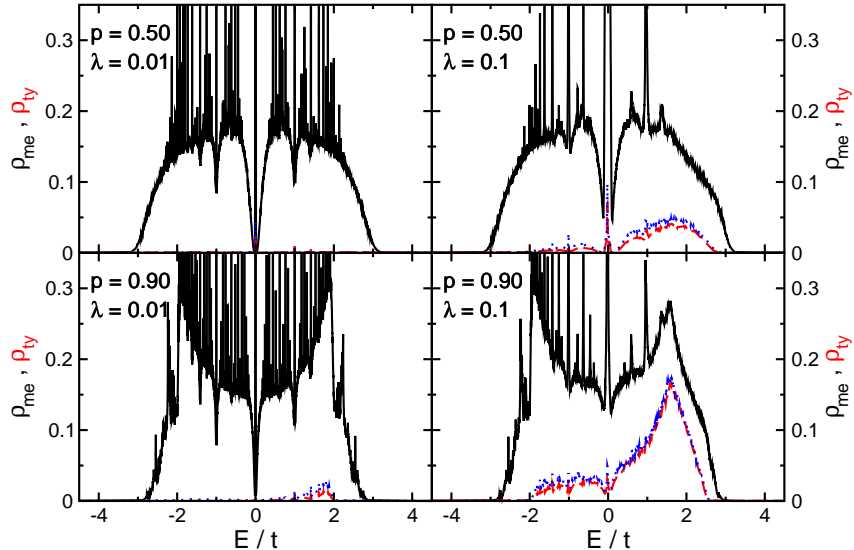}
   \caption{Mean (solid line) and typical (dashed line) DOS for the 
     RRN model. Results are shown for $N=500^2$ and $N\sigma=45$ for
     different anisotropy $p$ and leakage $\lambda$. For illustration of
     the finite size dependence, $\rho_{\text{ty}}$ for a $N=350^2$
     sites system is also included (dotted lines).
     The results base on the calculation
     of 1000 individual LDOS spectra for different cluster realizations and sites}
   \label{RRN_mety}
\end{figure}

First representative results for the RRN model are presented in Fig.~\ref{RRN_mety}.
In particular it shows the influence of finite leakage and anisotropy.
In contrast to the $2D$ percolation model, where the DOS spectra are completely
symmetric, the inclusion of next-nearest neighbor hopping
causes an asymmetry that grows with increasing $\lambda$.
For large $p$, the mean DOS is evocative of the $1D$ DOS, except for
the multitude of spikes, which we can attribute again to localized states on 
isolated islands. 
In the isotropic, low-leakage case, a vanishing $\rho_{\text{ty}}$
suggests that all states are localized.
Either increasing $p$ or $\lambda$ leads to a finite value of $\rho_{\text{ty}}$. 
But even at $p=0.90$ this effect is marginal for small 
$\lambda$, thus presumably no extended states exist.
Increasing the leakage results in a finite $\rho_{\text{ty}}$ for $E>0$ also
at $p=0.5$.
This feature becomes even more pronounced at high anisotropy, which indicates
the existence of extended states for these parameters.

\begin{figure}
  \centering 
  \includegraphics[width=0.95\linewidth]{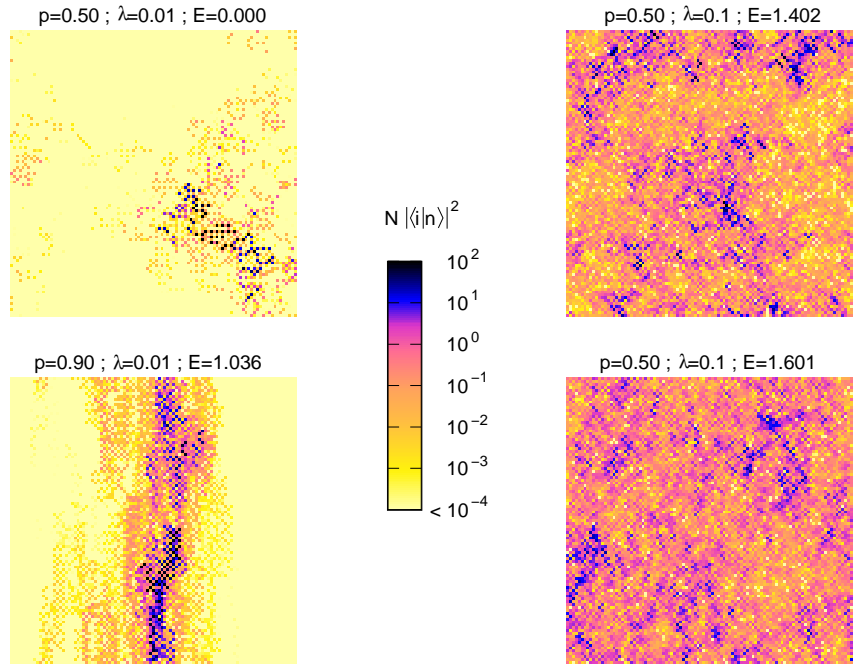}
   \caption{Characteristic eigenstates of the RRN model on a $N=100^2$ lattice at
   different anisotropies $p$, leakage rates $\lambda$, and energies $E$.
   Shown is the normalized occupation probability
   $N|\langle i| n\rangle|^2$}\label{RRN_waveampl}
\end{figure}

In Fig.~\ref{RRN_waveampl} we present some characteristic eigenstates of this model
for a fixed realization of disorder.
These results support the conclusions drawn from the typical DOS.
For the isotropic, low-leakage case we find a clearly localized state with internal 
checkerboard structure.
Increasing the anisotropy, many states are extended in one direction (cf. the
$1D$-like shape of $\rho_{\text{me}}$), while localized in the other one.
The leakage has a more drastic effect on the nature of the states, as
they are extended in some sense for both values of $p$.
The amplitudes on different lattice sites fluctuate over several orders 
of magnitude, however, explaining the reduced value of $\rho_{\text{ty}}$
as compared to $\rho_{\text{me}}$ (cf. Fig~\ref{RRN_mety}).

Due to the simplicity of the model, these results are surely not 
suitable to be compared to real experimental transport data, but can be
seen as a first step towards an at least qualitative understanding of 
the extraordinary transport properties in graphene.
In any case, also here the LD approach may serve as a reliable tool to 
discuss localization effects.

\subsection{Doped CMR Manganites}
\begin{figure}
  \centering 
  \includegraphics[width=0.95\linewidth]{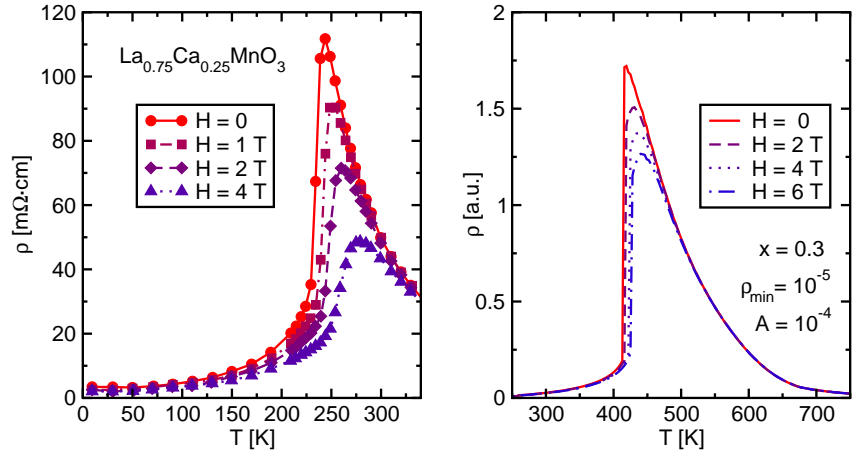}
   \caption{Experimental (left) and theoretical (right)
    results for the temperature- and magnetic field-dependence 
    of the resistivity in doped CMR manganites. Experimental
    data are taken from Ref.~\cite{SRBC95}. For a
    more detailed discussion see also~\cite{WLF03}
\label{CMR_Exp}}
\end{figure}

The transition from a metallic ferromagnetic (FM) low-temperature 
phase to an insulating paramagnetic high-temperature phase 
observed in some hole-doped manganese oxides (such as the perovskite family 
$\rm La_{1-x}[Sr,Ca]_xMnO_3$) is associated with an unusual 
dramatic change in their electronic and magnetic
properties. This  includes a spectacularly large negative magnetoresistive 
response to an applied magnetic field (see Fig.~\ref{CMR_Exp}, left panel),  
which might have important technological applications~\cite{JTMFRC94}.

Recent experiments indicate the coexistence of localized 
and itinerant charge carriers close to the metal-insulator transition 
in the FM phase of CMR manganites. Above $T_c$
the activated behaviour of the conductivity~\cite{WMG98} 
as well as the structure of the pair distribution function 
(PDF)~\cite{BDKNT96} indicate the formation
of small polarons, i.e., of almost localized carriers within a
surrounding lattice distortion. Interestingly these polarons continue
to exist in the metallic phase below $T_c$, 
merely their volume fraction is noticeable reduced. For the
coexistence of conducting and insulating regions within the metallic
phase different scenarios were discussed, 
which relate the metal-insulator transition to 
phase separation~\cite{DHM01} and percolative 
phenomena~\cite{GK98,MMFYD00}. In particular
microscopic imaging techniques, like scanning tunneling
spectroscopy~\cite{FFMTAM99,Beea02} or dark-field
imaging~\cite{UMCC99}, seem to support the latter idea. 

In previous work~\cite{WLF03} we addressed this problem 
theoretically. We proposed a phenomenological mixed-phase description 
which is based on the competition of a polaronic insulating phase and
a metallic, double-exchange driven ferromagnetic phase, 
whose volume fractions and carrier concentrations are
determined self-consistently by requiring equal pressure and 
chemical potential. In more detail, we assume that the resistivity of the
metallic component is proportional to the expression
\begin{equation}
  \rho_S[z] = \frac{g_S[z] - \gamma_S[z]^2}{\gamma_S[z]^2}\,,\\
\end{equation}
derived by Kubo and Ohata~\cite{KO72a}, which associates $\rho$ with
the fluctuation of the double-exchange matrix element caused by the
thermal spin disorder. Here $S=3/2$ is the localized 
spin formed by the $t_{2g}$ electrons of the manganese.
Both, 
\begin{align}
  g_S[z] &  = \frac{S
    B_S[z]}{(2S+1)^2}\left((2S+2)\coth\frac{(S+1)z}{S} -
  \coth\frac{z}{2S}\right)  + \frac{S+1}{2S+1}  \\
\gamma_{S}[z]&  = \frac{S+1}{2S+1}
  + \frac{S}{2S+1}\coth\left(\frac{S+1}{S}z\right)B_S[z] \;,
\end{align}
exhibit a magnetic field dependence, where $B_S[z] =
\frac{1}{2S}\big[(2S+1)\coth\tfrac{(2S+1)z}{2S}-\coth\tfrac{z}{2S}\big]$.
The resistivity of the insulating component is
assumed to match the resistivity of the high-temperature phase, which
in experiment is well fit by the activated hopping of
small-polarons~\cite{WMG98}. Hence, the resistivities of
the two components are given by,
\begin{equation}
  \rho^{(f)} = \frac{B}{x^{(f)}}
  \left(\rho_S[S(\lambda+\lambda^{\text{ext}})]+\rho_{\text{min}}\right)
  \quad\text{and}\quad
 \rho^{(p)}  = \frac{A}{\beta x^{(p)}}\ \rho_S[S\lambda^{\text{ext}}]
  \ \E^{-\beta \epsilon_p}\;,
\end{equation}
where $\epsilon_p$ is the polaron binding energy,
$\beta=1/k_BT$ the inverse temperature, $\lambda$ is the inner Weiss field,
$\lambda^{\text{ext}}$ ist the external magnetic field and 
the prefactors $A$ and $B$ as well as the cut-off
$\rho_{\text{min}}$ are free model parameters which could be
estimated from experimental data. Then, at a given doping level
$x$, i.e. chemical potential $\mu$, 
the resulting carrier concentrations $x^{(f)}$ 
and  $x^{(p)}$ in the coexisting regions define the 
two volume fractions by the equations
\begin{equation}
  x = p^{(f)} x^{(f)} + p^{(p)} x^{(p)}
  \quad\text{and}\quad
  p^{(f)} + p^{(p)} = 1\,,
\end{equation}
(see~\cite{WLF03}).
The resistivity of the whole sample, which may consist of an
inhomogeneous mixture of both components, is calculated on
the basis of a RRN. More precisely, we choose
nodes from a cubic lattice which belong to the metallic component with
probability $p^{(f)}$ and to the polaronic component with probability
$p^{(p)}$.  Each of these nodes, which represent macroscopic regions
of the sample, is connected to its neighbours with resistors of
magnitude $\rho^{(f)}$ or $\rho^{(p)}$, respectively.

Inserting the volume fractions and carrier concentrations from the
mixed-phase model  we obtain the resistivities 
shown in Fig.~\ref{CMR_Exp} (right panel). 
The jump-like behaviour of the resistivity
originates to a large degree from the changing volume fraction of the
metallic component, which can cross the percolation threshold.
However, the conductivity of the component itself as well as its
carrier concentration strongly affect $\rho$ for $T<T_c$. An external
magnetic field causes a reasonable suppression of $\rho$, i.e., a
noticeable negative magnetoresistance.  Compared to the real compounds
the calculated effect is a bit weaker. Nevertheless, in view of the
rather simple model for the conductivity the agreement is quite
satisfactory.  
More involved assumptions, e.g. an affinity to the formation 
of larger regions of the same type in the sense of correlated 
percolation~\cite{KK03} would naturally affect the resistivity of the 
system and its response to an external field.

\section{Conclusions}
\label{sec:conclusion}
In this tutorial we demonstrated the capability of the local 
distribution approach to the problem of quantum percolation. 
In disordered systems the local density of states 
(LDOS) emerges as a stochastic, random quantity.
It makes sense to take this stochastic character seriously and to
incorporate the distribution of the LDOS in a description of 
disorder. 
Employing the Kernel Polynomial Method we can resolve 
with very moderate computational costs the rich structures 
in the density of states originating from the irregular 
boundary of the spanning cluster. 

As for the standard Anderson localization and binary alloy 
problems the geometrically averaged (typical) density of states 
characterizes the LDOS distribution and may serve as a kind 
of order parameter differentiating between extended and 
localized states.
For both $2D$ and $3D$ quantum site percolation, 
our numerical data corroborate previous results in
favor of a quantum percolation threshold $p_q>p_c$ and a
fragmentation of the electronic spectrum into extended and 
localized states.
At the band center, so called anomalous localization is observed, 
which manifests itself in a checkerboard-like structure of the 
wave function.
Most notably, monitoring the spatial evolution of 
a wave packet in time for the $2D$ case, we find direct evidence
for 'dynamical' localization of an incident quantum particle 
at $p=0.65 > p_c$, while its wave function is 
spread over the percolated cluster for $p=0.9$.
This finding additionally supports the existence of a 
quantum percolation threshold. 
 
Without a doubt quantum percolation plays an important role
in the transport of several contemporary materials, such as 
$2D$ graphene or $3D$ manganese oxides.
To close the gap between the study of simple percolation models and a
realistic treatment of percolative transport in these rather 
complicated materials will certainly be a challenge of future research.

We thank A. Alvermann, F. X. Bronold, J. W. Kantelhardt, S. A. Trugman, A. Wei{\ss}e,
and G. Wellein for valuable discussions.
This work was supported by the DFG through the research program 
TR 24, the Competence Network for Technical/Scientific 
High-Performance Computing in Bavaria (KONWIHR), and the Leibniz 
Computing Center (LRZ) Munich.

\bibliography{schubert}
\bibliographystyle{spphys}

\printindex

\end{document}